# Giant magnetocaloric effect and hysteresis loss in Mn$_x$Fe$_{2-x}$P$_{0.5}$Si$_{0.5}$ (0.7 ≤ $x$ ≤ 1.2) microwires at ambient temperatures

Lin Luo[1], Hongxian Shen[1,*], Lunyong Zhang[1], Yongjiang Huang[1], Jianfei Sun[1,*], and Manh-Huong Phan[2,*]

[1] School of Materials Science and Engineering, Harbin Institute of Technology, Harbin 150001, China

[2] Department of Physics, University of South Florida, Tampa, FL 33620, USA

**Magnetocaloric microwires are very promising for energy-efficient magnetic refrigeration in micro electromechanical systems (MEMS) and nano electromechanical systems (NEMS). Creating microwires that exhibit large magnetocaloric effects around room temperature represents an important but challenging task. Here, we report a tunable giant magnetocaloric effect around room temperature in Mn$_x$Fe$_{2-x}$P$_{0.5}$Si$_{0.5}$ (0.7 ≤ $x$ ≤ 1.2) microwires by utilizing a melt-extraction technique paired with thermal treatment and chemical engineering. The isothermal magnetic entropy change ($\Delta S_{iso}$) and Curie temperature ($T_C$) can be tuned by adjusting the Mn/Fe ratio. The $T_C$ varies from 351 to 190 K as $x$ increases from 0.8 to 1.2. Among the compositions investigated, the $x$ = 0.9 sample shows the largest value of $\Delta S_{iso}$ = 18.3 J kg$^{-1}$ K$^{-1}$ for a field change of 5 T around 300 K. After subtracting magnetic hysteresis loss, a large refrigerant capacity of ~284.6 J kg$^{-1}$ is achieved. Our study paves a new pathway for the design of novel magnetocaloric microwires for active magnetic refrigeration at ambient temperatures.**

refrigeration

∗Corresponding authors: jfsun@hit.edu.cn (Jianfei Sun); hitshenhongxian@163.com (Hongxian Shen); phanm@usf.edu (Manh-Huong Phan)

**Introduction**

Solid-state magnetic refrigeration has attracted a great deal of attention due to its high energy efficiency and environmental friendliness.[1,2] In the past decades, many studies have focused on exploring magnetocaloric materials which are crucial components (also known as magnetic beds) in magnetic refrigerators.[3,4] However, current refrigerators operate at relatively low frequencies (up to a few ten Hertz), thus limiting the overall cooling efficiency.[5,6] Theoretical studies have suggested that the cooling efficiency of a magnetic refrigerator can be enhanced by optimizing the shape of the magnetocaloric materials used.[7,8] It has been predicted that a magnetic bed composed of magnetocaloric wires (e.g., Gd wires) yields an optimal device performance, as compared to their powder or laminate structures.[8] Owing to their increased surface-to-volume area, magnetocaloric wires also promise to promote a higher heat transfer between the magnetic refrigerant and a surrounding liquid. To realize these predictions, a large class of Gd-alloy microwires has been fabricated by the melt-extraction technique, and their magnetic and magnetocaloric properties have been extensively studied in recent years.[9-15] Due to their Curie temperature ($T_C$) regime being limited to the nitrogen temperature, however, the Gd-based microwires are only suited for magnetic refrigeration as nitrogen liquefaction.[11,15] Efforts to tune the magnetocaloric effect from nitrogen to room temperature have been made with little success.[16-19] While the Curie temperature appeared to be tunable, to some extent, by varying chemical compositions, the magnetic entropy change ($\Delta S_{iso}$) was found to decrease in Gd-based microwires with enhanced Curie temperatures.[11,15] This represents an important but challenging task in the field.

Among the magnetocaloric materials explored, Mn-Fe-P-Si alloys have been shown to exhibit a giant magnetocaloric effect around room temperature.[20,21] In addition to this outstanding property, the low cost of the material with no use of expensive or critical elements makes it one of the most promising refrigerants for large-scale cooling utilization. From the material engineering perspective, however, it is challenging to create Mn-Fe-P-Si alloys with high fractions of desirable $Fe_2P$ phase. Previous studies have shown that time-consuming heat-treatment (up to several days or months) at high temperature is needed to form a high volume fraction of the $Fe_2P$ phase, as well as to achieve compositionally homogeneous Mn-Fe-P-Si, using conventional methods.[22-24] This makes the total cost of producing the material much higher. Recently, rapid solidification such as melt-extraction technology has been employed to prepare Mn-Fe-P-Si microwires, which significantly reduces the heat treatment duration.[25] An elevated amount of the $Fe_2P$ phase was achieved, which could be increased following short-time (only 15 min) annealing. Compared to its as-quenched state, a two-fold increase in $\Delta S_{iso}$ was achieved in thermally treated Mn-Fe-P-Si microwires.[8] It is worth pointing out that the magnetic and magnetocaloric properties of $Fe_2P$-based materials are sensitive to changes in their composition and microstructures.[26,27] Mn-Fe-P-Si alloys prepared by different techniques were reported to yield uncontrollable chemical compositions resulting in low volume-fractions of the $Fe_2P$ phase.[28,29] Adjusting the Mn/Fe ratio appears to be crucial to tailor the composition and microstructure of Mn-Fe-P-Si, to improve its magnetocaloric response.

In this work, we aimed to harness melt-extraction technology with thermal treatment and chemical engineering to create Mn-Fe-P-Si microwires with large, tunable $\Delta S_{iso}$ and $T_C$ values at ambient temperatures. We systematically investigated the effect of varying the Mn/Fe ratio on the microstructure and magnetocaloric properties, as well as thermal and magnetic hysteresis losses of $Mn_xFe_{2-x}P_{0.5}Si_{0.5}$ (0.7≤$x$≤1.2) microwires. We have found that the $x$ = 0.9 sample possesses the largest $\Delta S_{iso}$ value of ~18.3 J kg$^{-1}$ K$^{-1}$ for a field change of 5 around 300 K. A high refrigerant capacity of ~284.6 J kg$^{-1}$ after subtracting the magnetic hysteresis loss is also achieved. Our study provides an effective approach for designing microwires with novel magnetocaloric properties for application in solid-state cooling systems at ambient temperatures.

**Results and Discussion**

*Microstructural characterization*

The XRD patterns of $Mn_xFe_{2-x}P_{0.5}Si_{0.5}$ (0.7≤$x$≤1.2) microwires at room temperature revealed a hexagonal Fe$_2$P-type structure (space group $P\bar{6}2\,m$) in addition to an impurity (MnFe)$_3$Si (space group $Fm\bar{3}m$) phase, as shown in Fig. 1(a). Tegus *et al.* [31] analyzed the XRD patterns of Fe$_2$P-type Mn-Fe-P-As compounds and concluded that the XRD patterns could be indexed with the Fe$_2$P structure type, and the Fe$_2$P-type compounds underwent a paramagnetic to ferromagnetic transition as the overlapping (211) and (002) peaks became separated. Wang *et al.* also have suggested that the well-separating (211) and (002) peaks bear the ferromagnetic (FM) characteristic of the Fe$_2$P phase, while the overlapping (211) and (002) peaks imply the paramagnetic (PM) nature of the Fe$_2$P phase.[32] According to the XRD patterns collected at ambient temperature (Fig. 1a), it can be inferred that the Fe$_2$P phase is FM for $x=0.7$ but PM for $x\geq0.9$, whereas two phases coexist for $x=0.8$.

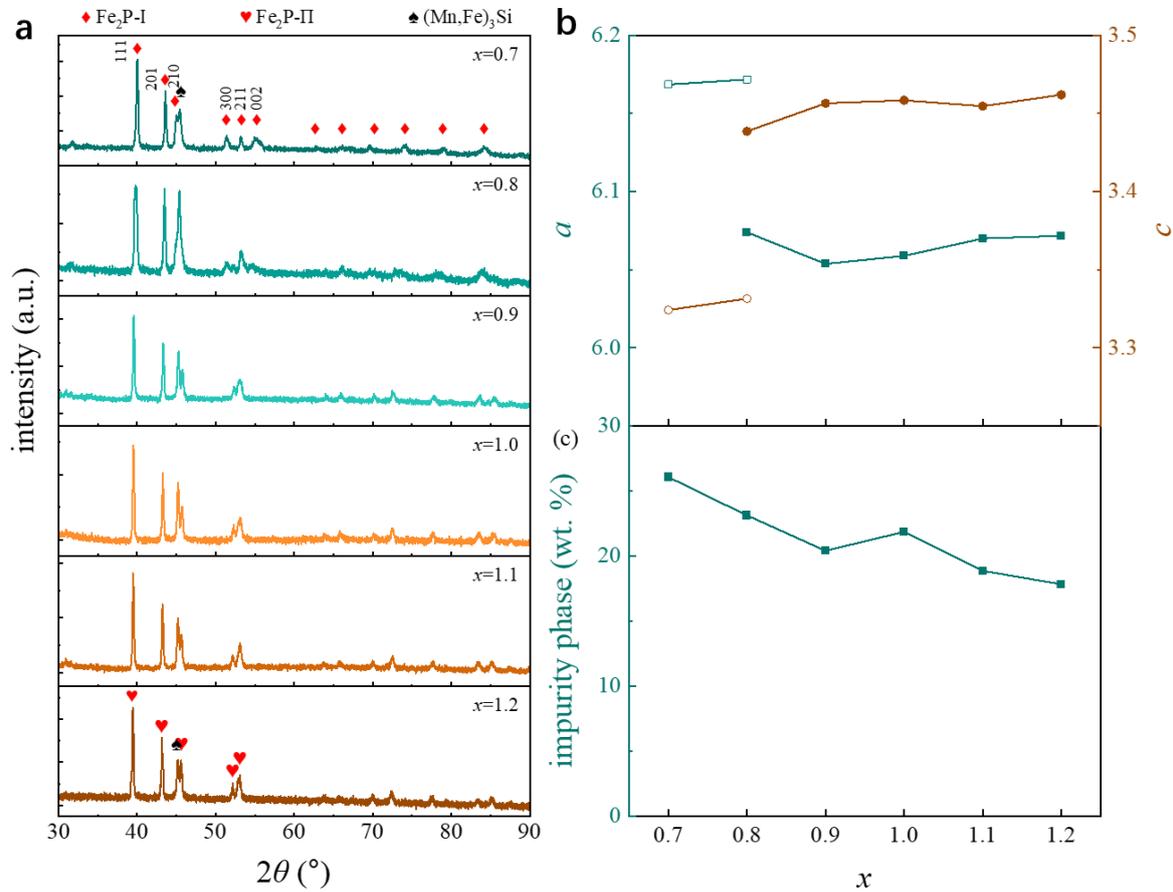

Fig. 1. (a) The room temperature XRD patterns and (b, c) Rietveld refinement results of the lattice constants and impurity phase fraction of Mn$_x$Fe$_{2-x}$P$_{0.5}$Si$_{0.5}$ microwires as the Mn content is varied. Lattice parameters ($a$ and $c$) are in Å.

The Rietveld refinement results, as shown in Fig.1(b) and (c), provide further support to the conclusion drawn by the XRD patterns. As shown in Fig. 2(a), the FM Fe$_2$P phase for $x=0.7$ shows a larger lattice $a$ and a smaller $c$ than the PM Fe$_2$P phase for $x\geq0.9$. Both the PM and FM phases coexist for $x=0.8$. In addition, the amount of impurity phase (Mn,Fe)$_3$Si decreased as the Mn content $x$ increased. The decrease of the (Mn,Fe)$_3$Si phase could be attributed to the decrease of Fe content.

The SEM images (Backscatter Electron mode) and the EDS analysis results of the cross section for a representative sample $Mn_{1.0}Fe_{1.0}P_{0.5}Si_{0.5}$ are shown in Fig. 2. The impurity visible in light gray zone is observed as a long strip shape or oval shape. The EDS maps clearly show that P is rich in the main phase and almost negligible in the impurity phase. Furthermore, the EDS line scans indicate that more Mn atoms entered the main phase, while more Fe and Si atoms entered the impurity phase. The compositions of these two phases in $Mn_xFe_{2-x}P_{0.5}Si_{0.5}$ microwires were analyzed by EDS, which are listed in Table 1. The above results further specify that the main phase is 2:1 and the impurity phase is 3:1. Accordingly, it is understandable that the composition of the $Fe_2P$ phase deviates from its nominal composition due to the presence of the $(Mn,Fe)_3Si$ phase.

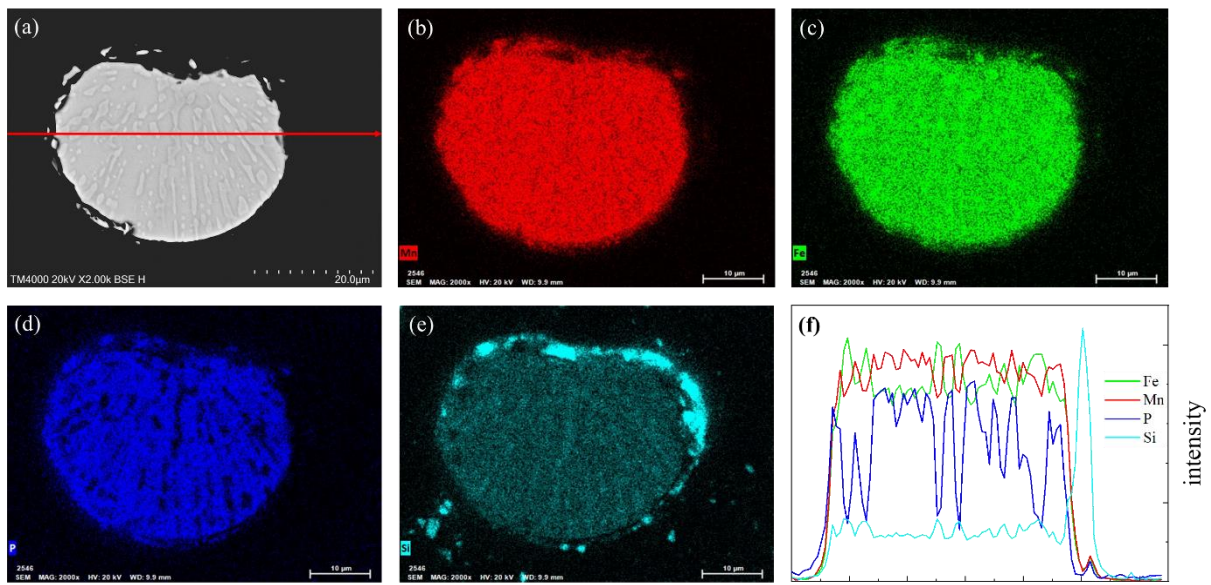

Fig. 2(a) The SEM image, (b-e) EDS color maps, and (f) line scans along the direction of the red arrow of the cross section for the $Mn_{1.0}Fe_{1.0}P_{0.5}Si_{0.5}$ microwire.

Table 1. The compositions of the two phases in $Mn_xFe_{2-x}P_{0.5}Si_{0.5}$ microwires

| x | Main phase | | Impurity | |
|---|---|---|---|---|
| | composition | M/NM | composition | M/NM |
| 0.7 | $Mn_{27.5}Fe_{40.8}P_{19.1}Si_{12.7}$ | 2.15 | $Mn_{19.6}Fe_{56.1}P_{3.7}Si_{20.5}$ | 3.14 |
| 0.8 | $Mn_{30.3}Fe_{38.2}P_{18.1}Si_{13.4}$ | 2.18 | $Mn_{23.0}Fe_{51.6}P_{4.6}Si_{20.8}$ | 2.94 |
| 0.9 | $Mn_{32.9}Fe_{35.2}P_{19.0}Si_{12.9}$ | 2.14 | $Mn_{27.1}Fe_{48.0}P_{4.5}Si_{20.5}$ | 3.01 |
| 1.0 | $Mn_{36.5}Fe_{30.7}P_{20.0}Si_{12.8}$ | 2.05 | $Mn_{30.8}Fe_{43.9}P_{4.9}Si_{20.3}$ | 2.96 |
| 1.1 | $Mn_{39.7}Fe_{28.5}P_{18.5}Si_{13.4}$ | 2.14 | $Mn_{37.0}Fe_{35.6}P_{8.3}Si_{19.1}$ | 2.64 |
| 1.2 | $Mn_{42.3}Fe_{25.4}P_{19.6}Si_{12.7}$ | 2.10 | $Mn_{39.0}Fe_{36.8}P_{2.8}Si_{21.5}$ | 3.12 |

Figure 3 shows the phase map and IPF map obtained by EBSD measurements of the $Mn_{1.0}Fe_{1.0}P_{0.5}Si_{0.5}$ microwire. The phase distribution can be seen clearly from the phase map, as shown in Fig. 3(a). The (Mn, Fe)$_3$Si phase was dispersed throughout the cross-section, no segregation occurred. Fig. 3(b) displays the micron-size grain without preferential orientation.

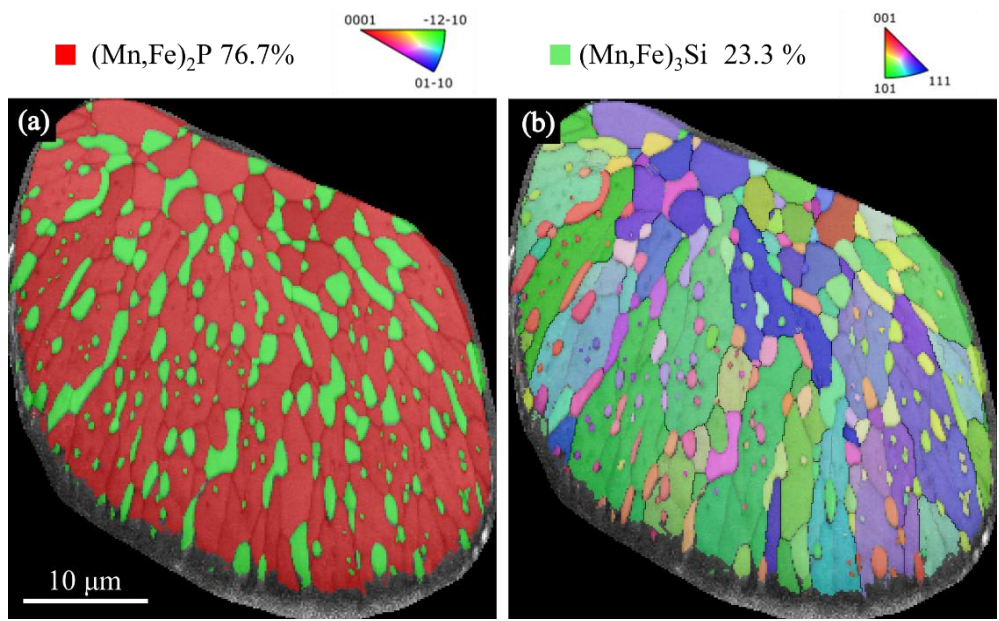

Fig. 3. EBSD analysis: (a) Phase map and (b) IPF map of the $Mn_{1.0}Fe_{1.0}P_{0.5}Si_{0.5}$ microwire.

The high-angle annular dark field (HAADF) image of the $Mn_{1.0}Fe_{1.0}P_{0.5}Si_{0.5}$ microwire, as shown in Fig. 4(a), confirmed that the crystallite size is in the micron range. Figs. 4(b) and (d) show the high-resolution TEM (HRTEM) images of grain A and B in Fig. 4(a). The grain A and B are confirmed in the hexagonal Fe$_2$P-type structure (space group $P\bar{6}2m$) and cubic

Fe$_3$Si-type structure (space group $Fm\bar{3}m$) respectively by performing the fast Fourier transform (FFT) of the corresponding HRTEM images, as shown in Fig. 4(c) and (e). The zone axis is identified to be [001] for both grains. Figs 4(f-i) show the corresponding energy dispersive spectroscopy (EDS) maps for Mn, Fe, P, and Si of Fig. 4(a), confirming that more Fe and Si atoms are present with negligible P atoms in the (Mn,Fe)$_3$Si phase.

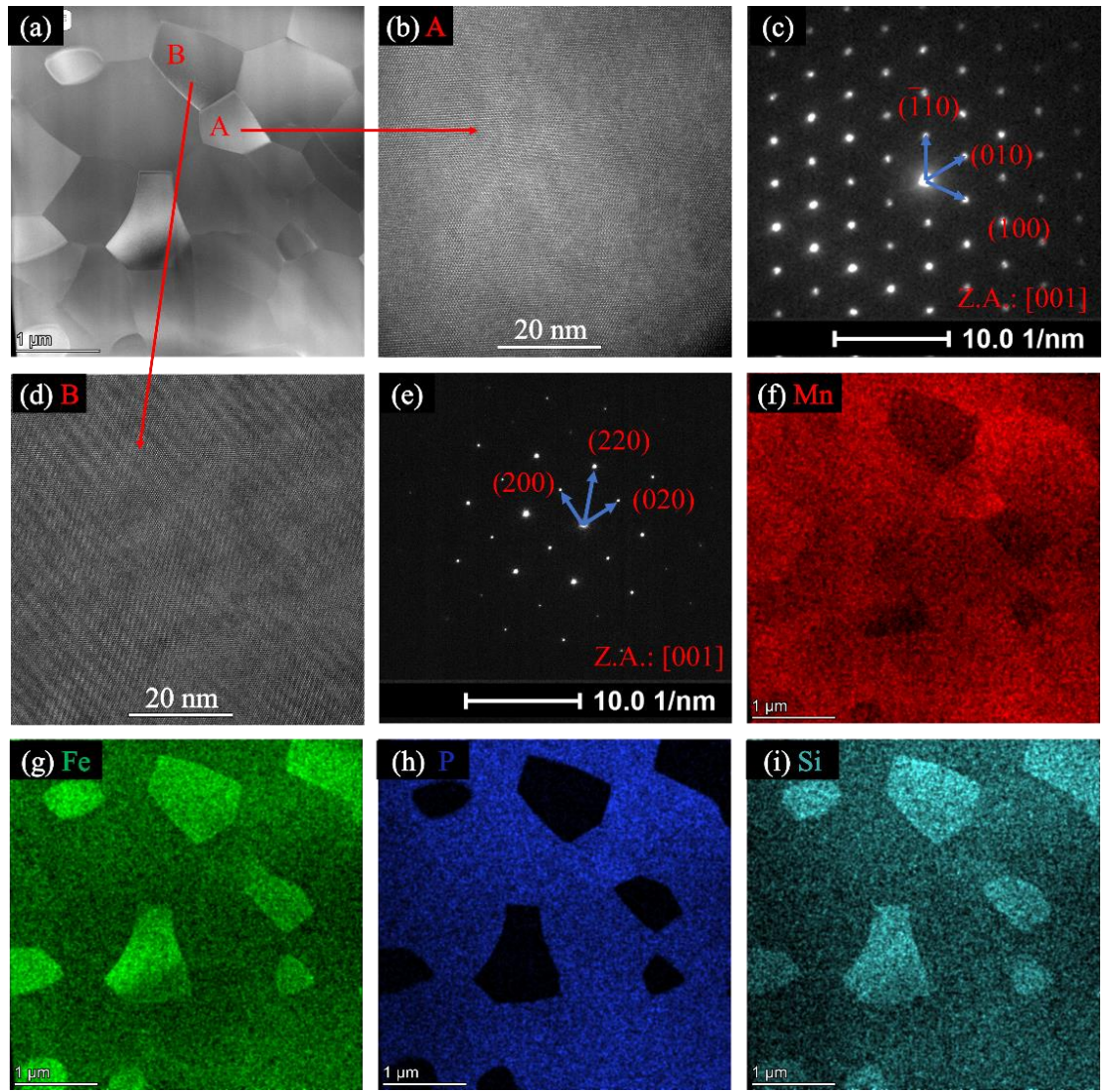

Fig. 4. (a) The high-angle annular dark field (HAADF) image, HRTEM images and the FFT for the grains (b, c) A and (d, e) B, (f-i) TEM-EDS maps of the Mn$_{1.0}$Fe$_{1.0}$P$_{0.5}$Si$_{0.5}$ microwire.

*Magnetic and magnetocaloric properties*

A PPMS paired with the vibrating sample magnetometer (VSM) option was utilized to characterize the magnetic properties of the microwires. The temperature-dependent magnetization (*M-T*) curves, the Mn-content dependence of the Curie temperature ($T_C$) and thermal hysteresis ($T_{hys}$), for $Mn_xFe_{2-x}P_{0.5}Si_{0.5}$ microwires under a magnetic field of 0.05 T are displayed in Fig. 5. The deduced magnetic parameters are listed in Table 2. All the *M-T* curves show an FM-PM transition, with thermal hysteresis observed between field-warming (FW) and field-cooling (FC) curves. For *x*=0.7, the Curie temperature exceeds 400 K, which is above the experimental temperature range available. The $T_C$ was found to decrease from 351 to 190 K as the Mn content *x* increased from 0.8 to 1.2. The decrease in $T_C$ could be mainly caused by the changes in the Mn/Fe ratio. In addition, the $T_{hys}$ increased from 14.5 to 22 K as the Mn content *x* increased.

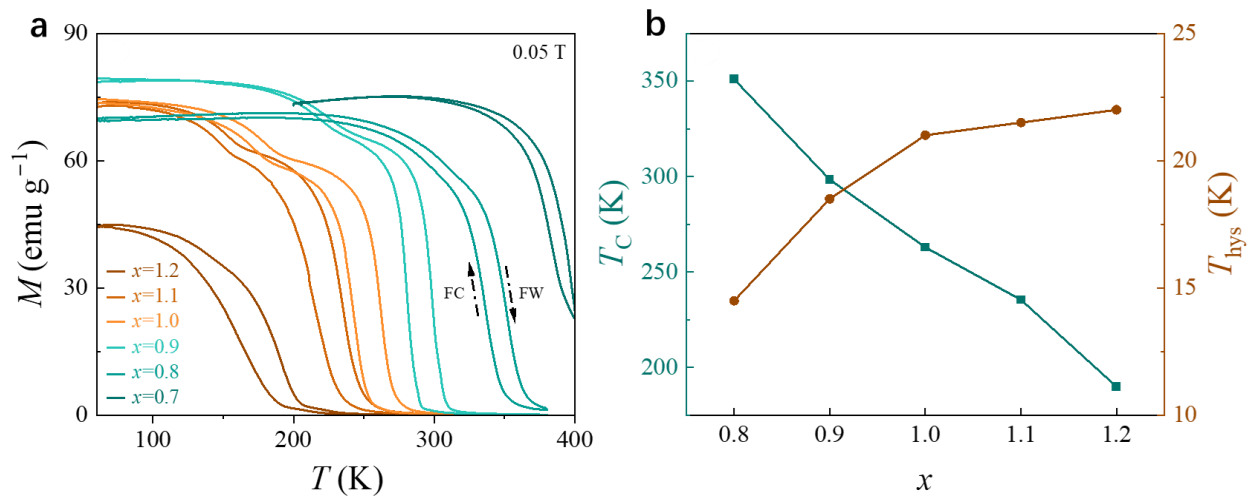

Fig. 5 (a) The temperature dependence of magnetization under FW and FC measurement protocols, (b) the Curie temperature and thermal hysteresis of $Mn_xFe_{2-x}P_{0.5}Si_{0.5}$ microwires.

Table 2. The magnetic and magnetocaloric properties of $Mn_xFe_{2-x}P_{0.5}Si_{0.5}$ microwires

| $x$ | $T_C$ (K) | $T_{hys}$ (K) | $W_y^{peak}$ (J kg$^{-1}$) | $\Delta S_{iso}^{peak}$ (J kg$^{-1}$ K$^{-1}$) | $RC$ (J kg$^{-1}$) | $RCE$ (J kg$^{-1}$) |
|---|---|---|---|---|---|---|
| 0.7 | > 400 | - | - | - | - | - |
| 0.8 | 351 | 14.5 | 19.6 | 12.0 | 293.7 | 279.2 |
| 0.9 | 298.5 | 18.5 | 60.7 | 18.3 | 331.1 | 284.6 |
| 1.0 | 263 | 21 | 58.4 | 15.8 | 300 | 257.1 |
| 1.1 | 235.5 | 21.5 | 28.7 | 10.9 | 280.9 | 262.3 |
| 1.2 | 190 | 22 | 28.9 | 7.9 | 288.4 | 267.5 |

Figure 6a displays the isothermal magnetization curves of $Mn_{1.0}Fe_{1.0}P_{0.5}Si_{0.5}$ microwires under applied magnetic fields ranging from 0-5 T. The non-coincidence of magnetization and demagnetization $M$-$H$ curves in Fig. 6a indicates the existence of magnetic hysteresis around the $T_C$, arising from the type of the first-order phase transition that occurs in the microwires. The magnetic hysteresis loss ($W_y$) is defined as the area between the magnetization and demagnetization curves.[33] Accordingly, the temperature dependence of $W_y$ for $Mn_xFe_{2-x}P_{0.5}Si_{0.5}$ microwires is estimated, and the results are shown in Fig. 6b. Larger values of $W_y$ were found for $x$ = 0.9 and 1.0, in which the Mn/Fe ratio of the Fe$_2$P phase trends towards 1, while the other samples show smaller magnetic hysteresis losses.

The isothermal magnetic entropy change ($\Delta S_{iso}$), which is one of the important parameters for evaluating the magnetocaloric effect of a magnetic material, was calculated using the isothermal magnetization data (e.g., Fig. 4a), via Maxwell equation, i.e., eq. (1):

$$\Delta S_{iso} = \mu_0 \int_0^H \left(\frac{\partial M}{\partial T}\right)_{H'} dH'. \tag{1}$$

Figure 6c shows a plot of variation of isothermal magnetic entropy change with temperature for $Mn_xFe_{2-x}P_{0.5}Si_{0.5}$ microwires for a field change of 5 T. It can be seen in this figure that the peak value of $\Delta S_{iso}$ increased from 7.9 to 18.3 J kg$^{-1}$ K$^{-1}$ with decreasing $x$ from 1.2 to 0.9, maximized at $x=0.9$, and then decreased to 12.0 as $x$ continued to decrease to 0.8. This changing trend in $\Delta S_{iso}$ can be attributed to the change in the Mn/Fe ratio and hence the fraction of the Fe$_2$P phase. Interestingly, the largest value of $\Delta S_{iso}$ is achieved near room temperature for the $x=0.9$ composition.

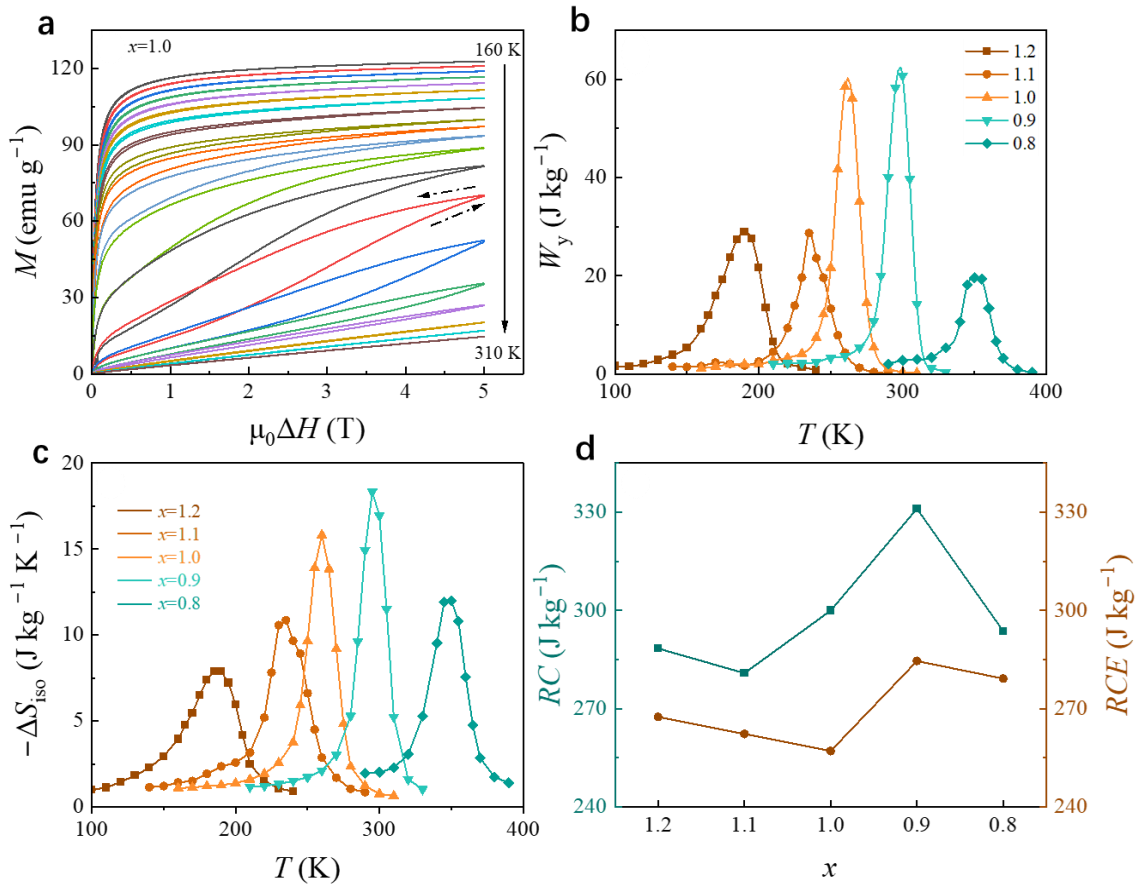

Fig. 6 (a) The isothermal magnetization $M$-$H$ curves of Mn$_{1.0}$Fe$_{1.0}$P$_{0.5}$Si$_{0.5}$ ($x=1.0$) microwires, (b) the magnetic hysteresis loss ($W_y$), (c) the isothermal magnetic entropy change ($\Delta S_{iso}$), (d) refrigerant capacity ($RC$) and effective refrigerant capacity ($RCE$) of Mn$_x$Fe$_{2-x}$P$_{0.5}$Si$_{0.5}$ microwires for a field change of 5 T.

In addition to the isothermal magnetic entropy change, the refrigerant capacity ($RC$) is another important parameter for the evaluation of the cooling efficiency of magnetocaloric materials. The $RC$ is determined by the integral area in the temperature range of full width at half maximum under the $\Delta S_{iso}$–$T$ curves[12], i.e., Eq. (2):

$$RC = \int_{T_1}^{T_2} -\Delta S_{iso}(T)dT, \qquad (2)$$

where $T_1$ and $T_2$ are the corresponding temperatures of half maximum of isothermal magnetic entropy change, and $T_1 \leqslant T_2$. Taking the magnetic hysteresis losses into account, the $RC$ values after subtracting the average magnetic hysteresis loss ($\Delta W_y$) are referred as to the effective refrigerant capacity ($RCE$) values. The $\Delta W_y$ is determined by the averaging integral area under the $W_y$-$T$ curves using the same integration limit as $RC$ [28,33], i.e., Eq. (3):

$$\Delta w_y = \frac{\int_{T_1}^{T_2} W_y(T)dT}{T_2 - T_1} \qquad (3)$$

The $RC$ and $RCE$ values obtained for $Mn_xFe_{2-x}P_{0.5}Si_{0.5}$ microwires for a field change of 5 T are plotted versus the Mn content, as shown in Fig. 6(d). By varying the Mn content, the $RC$ shows a similar trend as $\Delta S_{iso}$ except for $x$=1.2. In comparison with the $x$=1.1 sample, the $x$=1.2 sample shows the larger $RC$, and the latter possesses a smaller $\Delta S_{iso}$ value than the former. This could be attributed to the broader FM-PM transition for the $x$=1.2 sample as compared to the $x$=1.1 sample. The largest $RC$ and $RCE$ values (331.1 and 284.6 J kg$^{-1}$, respectively) are also achieved for the $x$=0.9 sample. The magnetic and magnetocaloric properties of the $Mn_xFe_{2-x}P_{0.5}Si_{0.5}$ ($x$=0.9) microwires (MW-1 h) and their bulk counterparts (Figs. S1 and S1 in Supporting Information) are listed in Table 3. It can be seen in this table that one bulk counterpart under the same heat-treatment (B-1 h) showed a larger $\Delta S_{iso}$ but a similar $RCE$. Another bulk counterpart, subjected to the heat treatment for 4 days (B-4 d), showed a decreased $RCE$ despite an increase in $\Delta S_{iso}$, due to its increased hysteresis loss. These values of $\Delta S_{iso}$ are much larger than those achieved for Gd-based microwires with much lower $T_C$ values

[9-15], and for the magnetocaloric materials with $T_\text{C}$ around 300 K (see Table 3). These results indicate that the Mn$_x$Fe$_{2-x}$P$_{0.5}$Si$_{0.5}$ ($x$=0.9) microwires are very compelling candidates for active magnetic refrigeration in the room temperature regime.

Table 3. The magnetic and magnetocaloric properties of magnetocaloric materials

| Materials | $T_\text{C}$ (K) | $T_\text{hys}$ (K) | $W_y^{peak}$ (J kg$^{-1}$) | $\Delta S_{iso}^{peak}$ (J kg$^{-1}$ K$^{-1}$) | RC (J kg$^{-1}$) | RCE (J kg$^{-1}$) | Ref. |
|---|---|---|---|---|---|---|---|
| MW-1 h | 298.5 | 18.5 | 60.7 | 18.3 | 331.1 | 284.6 | This work |
| B-1 h | 350 | 13 | 18.1 | 6.6 | 305.9 | 293.2 | This work |
| B-4 d | 319 | 33.5 | 155.6 | 22.4 | 360.1 | 242.3 | This work |
| Mn$_{1.20}$Fe$_{0.75}$P$_{0.45}$Si$_{0.55}$ | 301.5 | 1 | - | 11.9 | - | - | [34] |
| MnFeP$_{0.5}$Si$_{0.5}$ | 282 | 40 | - | 18.1 | | | [35] |
| Gd | 290 | - | | 9.8 | | | [36] |
| LaFe$_{11.5}$Al$_{1.5}$H$_{1.3}$ | 295 | - | | 12.3 | | | [37] |
| Ni$_{52.9}$Mn$_{22.4}$Ga$_{24.7}$ | 305 | | | 8.6 | | | [38] |

**Conclusions**

We have successfully created a novel class of Mn$_x$Fe$_{2-x}$P$_{0.5}$Si$_{0.5}$ microwires with tunable giant magnetocaloric properties around room temperature, by harnessing the combined advantages of melt-extraction, thermal treatment, and chemical engineering. The Mn$_x$Fe$_{2-x}$P$_{0.5}$Si$_{0.5}$ microwires crystallize into an Fe$_2$P-type structure as a majority phase, with (Mn,Fe)$_3$Si present as an impurity phase whose volume fraction decreases with increasing Mn content. By varying the Mn content $x$ from 0.8 to 1.2, the Curie temperature ($T_\text{C}$) can be adjusted from 351 to 190 K while the thermal hysteresis ($T_\text{hys}$) can be altered between 15 and 22 K. Among the compositions investigated, the largest values of $\Delta S_\text{iso}$, RC, and RCE (18.3 J kg$^{-1}$ K$^{-1}$, 331.1 and 284.6 J kg$^{-1}$, respectively) are achieved for $x$=0.9 at ambient temperature. The adjustability of the microstructural and magnetic properties combined with a short-time heat treatment identifies melt-extracted Mn-Fe-P-Si microwires as a strong candidate for active magnetic refrigeration at room temperature.

**Methods**

*Fabrication*

$Mn_xFe_{2-x}P_{0.5}Si_{0.5}$ ($0.7 \leqslant x \leqslant 1.2$) microwires of ~50 μm diameter were prepared by melt-extraction, followed by subsequent heat-treatment. First, an 8 mm diameter rod was prepared by arc-melting and casting. The raw materials were crystals of Mn (99.5 %), Fe (99.99 %), FeP chunks (98 %), and Si (99.999 %). An excess of 5 *wt*.% of Mn was added to compensate for the loss during arc-melting. Then, the rod was placed in a homemade melt-extractor system to be transformed into microwires.[30] Finally, the as-prepared microwires were sealed in a quartz tube under 30 KPa Ar atmosphere and annealed at 1323 K for 1 h and then water quenched.

*Structural and magnetic characterization*

The microwires were ground into powders for X-ray diffraction (XRD) characterization, and the XRD patterns of the powdered samples were collected from 30 to 90° on a PANalytical X'Pert Pro diffractometer (Empyrean) with Cu-Kα radiation at room temperature. Rietveld refinement was performed using Fullprof program for obtaining the structural parameters and phase fractions. Microstructural and compositional information were obtained using a scanning electron microscope (SEM, TM4000) with energy dispersive spectrometer (EDS). Electron back-scatter diffraction (EBSD) and transmission electron microscopy (TEM) measurements were also performed on a field emission scanning electron microscope (Gemini560) and a FEI Talos F200X system, respectively. Magnetic measurements were carried out on a Quantum Design Physical Property Measurement System (PPMS-16 T).

**Data availability.** The data that support the findings of this study are available from the corresponding author on a reasonable request.

**Acknowledgments**

Research at HIT was funded by the National Natural Science Foundation of China (NSFC, Nos. 51871124), grant PID2019-105720RB-I00 funded by MCIN/AEI /10.13039/501100011033. M.H.P acknowledges support from the University of South Florida under Grant No. 125300. Noah Schulz is acknowledged for proofreading the manuscript.

**Author contributions**

L.L. and H.X.S. initiated the research. L.L., H.X.S., L.Y.Z., Y.J.H., J.F.S. participated in the structural and magnetic experiments. L.L., H.X.S. and M.H.P. analyzed the data. L.L., H.X.S. and M.H.P. prepared the first draft of the manuscript and all the co-authors contributed to the final version of the manuscript. J.F.S. led the project.

**Competing interests:** The authors declare no conflict of interest.